\documentclass[aps,prl,twocolumn,floats,showpacs]{revtex4}
\usepackage{epsfig}
\usepackage{amsmath}
\begin{document}
\title{Explicit and Hidden Symmetries in Quantum Dots and
Quantum Ladders}
\author{K. Kikoin, Y. Avishai,}
\address{Ben-Gurion University of the Negev, Beer-Sheva 84105,
Israel}
\author{M.N. Kiselev}
\address{Institut f\"ur Theoretische Physik, Universit\"at
W\"urzburg, D-97074 W\"urzburg, Germany}
\begin{abstract}
The concept of dynamical hidden symmetries in the physics of
electron tunneling through composite quantum dots (CQD) and
quantum ladders (QL) is developed and elucidated. Quite generally,
dynamical symmetries are realizable in the space of low energy
excited states in a given charge sector of nanoobjects, which
involve spin variables and/or electron-hole pairs. While spin
multiplets in an individual rung of a QL or in an isolated CQD
 form a representation space of the usual
rotation group, this $SU(2)$ symmetry is broken due to spin
transfer (in QL) electron cotunneling through CQD. Dynamical
symmetries in the space of spin multiplets are then unravelled in
these processes. The corresponding symmetry groups are described
by $SO(n)$ or $SU(n)$ depending on the origin of rotation group
symmetry breaking. The effective spin Hamiltonians of QL and CQD
are derived and expressed in terms of the pertinent group
generators. We employ fermionization procedure for analyzing the
physical content of these dynamical symmetries, including Kondo
tunneling through CQD and Haldane gap formation in QL.
\end{abstract}
\pacs{72.10.-d, 72.15.-v, 73.63.-b} \maketitle
\section{Introduction}
Symmetry considerations play a central role in the physics of
low-dimensional systems, and serves as a key for understanding
their peculiar properties \cite{Tsv}. It predetermines their
thermodynamics, response to external fields, transport properties,
phase diagrams, etc. In many cases, formulating the physics of
strongly interacting electrons in low dimensional systems
(especially nano-objects) should be constructed by employing group
theoretical concepts (among them non-commutative algebras), whose
specific structure have direct consequences for observable
physical properties. It appears that not only the symmetry of a
given Hamiltonian but also the {\it dynamical symmetry} of
low-energy excitations is relevant.

In order to clarify the above statements, let us consider a system
with Hamiltonian ${\cal H}_0$ whose eigenstates
$|\Lambda\rangle=|M\mu\rangle$ form a basis for an irreducible
representation of some Lie group $G$ ($\mu$ enumerates the lines
of this representation). It is convenient to express the
generators of Lie algebras via Hubbard operators
$X^{\Lambda\Lambda'}=|\Lambda\rangle\langle\Lambda'|$. Then the
Hamiltonian under consideration is expressed in terms of diagonal
Hubbard operators
\begin{equation}
{\cal H}_0=\sum_{\Lambda=M\mu}E_{\Lambda} |\Lambda\rangle\langle
\Lambda| =\sum_{\Lambda}E_{M} X^{\Lambda \Lambda} ~, \label{HX}
\end{equation}
so that
\begin{equation}\label{comm}
[X^{\Lambda\Lambda^{\prime}},{\cal H}_0] =
-(E_{M}-E_{M^{\prime}})X^{\Lambda \Lambda^{\prime}}.
\end{equation}
The symmetry group of the Hamiltonian is then generated by the
operators $X^{M\mu,M\mu'}$, which commute with ${\cal H}_0$,
whereas the dynamical symmetry of ${\cal H}_0$ is generated by the
whole set of operators $\{X\}$. This dynamical symmetry may be
revealed, when ${\cal H}_0$ describes a quantum object, which is
part of larger system with Hamiltonian ${\cal H}$, and its
symmetry is violated by interaction with this environment. If the
interaction scale is characterized by some energy ${\cal E}$, then
the dynamical symmetry is determined by transitions between those
states from the manifold $E_\Lambda$, which fall into the interval
${\cal E}$. We divide the Hubbard operators acting within this
low-energy interval into subsets $\{S\}$ and $\{R\}$. Here the
$S$-operators generate the symmetry group $G$, whereas the $S$-
and $R$-operators together generate the dynamical group $D$. In
this paper we study spin properties of quantum dots and quantum
ladders, so the group $G$ is in fact $SU(2)$, namely, the group of
spin angular momentum. It will be shown that the dynamical
symmetry of this object is that of the $SO(n)$ group. We will
construct the corresponding algebras by means of Hubbard
operators, rewrite the corresponding Hamiltonians ${\cal H}$ in
terms of the group generators, discuss the possible ways of
fermionization of these Hamiltonians and consider some specific
properties of quantum dots and quantum ladders possesing these
symmetries.
\section{From spin rotator to Kondo tunneling}
The symmetry of spin rotator is an intrinsic property of many
low-dimensional spin systems. As was shown in \cite{KA01}, this
symmetry predetermines the low-energy dynamics of zero-dimensional
quantum dots with even occupation in tunneling contact with
metallic reservoirs. Let us consider a double quantum dot
(DQD)occupied by two electrons in a neutral state in a T-shaped
parallel geometry (Fig.1) as a representative example.
\begin{figure}[h]
\centerline{\epsfig{figure=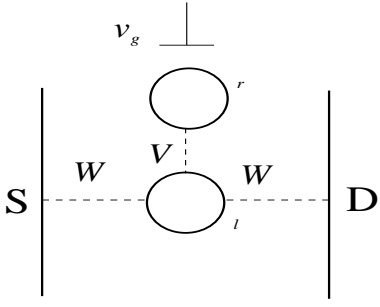,width=50mm,height=40mm,angle=0}}
\caption{Parallel Double Quantum Dots in contact with source (S)
and drain (D) metallic leads. $V$ and $W$ are tunneling coupling
constants, $v_g$ is a gate voltage.} \label{fig1}
\end{figure}
In this geometry, two valleys of DQD are coupled by tunneling $V$.
In a limit of strong Coulomb blockade $Q$, such that $V\ll Q$, the
energy spectrum of an isolated DQD consists of a ground state
singlet with energy $E_S$, a spin triplet with the energy $E_T$
separated by an exchange gap $\delta=2V^2/Q$ above $E_S$ and two
charge transfer excitons with large excitation energies $\sim Q$,
the charging energy for a given well of DQD. Thus the indices
$\Lambda$ in the Hamiltonian ${\cal H}_0$ (\ref{HX}) stand for
$\Lambda=S, T\mu$ with $\mu=1,0,\bar{1}$ denotes the three
projections of the spin $S=1$.

The dynamical symmetry of the $\{S,T\}$ manifold is that
appropriate for the $SO(4)$ group. Two vectors generating this
group are constructed by means of Hubbard operators (\ref{comm})
in the following way:
\begin{eqnarray}
S^+ & = & \sqrt{2}\left(X^{10}+X^{0-1}\right),~
S^-  =  \sqrt{2}\left(X^{01}+X^{-10}\right),\nonumber \\
S_z &  = &  X^{11}-X^{-1-1}. \label{m.1}
\end{eqnarray}
\begin{eqnarray}
R^+ & = & \sqrt{2}\left(X^{1S}-X^{S-1}\right), R^-  =
\sqrt{2}\left(X^{S1}-X^{-1S}\right),
\nonumber\\
R_z & = & -\left(X^{0S}+X^{S0}\right). \label{SP}
\end{eqnarray}
The first one, ${\bf S}$ is the conventional spin 1 operator,
while the second vector is the $R$-operator describing $S/T$
transitions. The spin algebra is $o_4$, which is characterized by
the commutation relations
\begin{eqnarray}
&&[S_\alpha,S_\beta]  =
ie_{\alpha\beta\gamma}S_\gamma,~[R_\alpha,R_\beta]=ie_{\alpha\beta\gamma}S_\gamma,~
[R_\alpha,S_\beta]=ie_{\alpha\beta\gamma}R_\gamma . \label{comm1}
\end{eqnarray}
($\alpha,\beta,\gamma$ are Cartesian coordinates,
$e_{\alpha\beta\gamma}$ is a Levi-Civita tensor). These vectors
are orthogonal, ${\bf S\cdot R} = 0,$, the Casimir operator is
${\bf S}^2+ {\bf R}^2 =3.$

A gate voltage $v_g$ applied to DQD, makes the level positions
essentially asymmetric, and the charging energy $Q$ may be nearly
compensated for at least one of the charge transfer singlet
excitons (say, the right one, $\Lambda=E_r$). In this case we
encounter a "Coulomb resonance" excitations, where the spin
singlet and charge transfer exciton (also a singlet!) are strongly
intermixed, but the spin triplet is  untouched by this resonance
tunneling. This means that the corresponding manifold is
$\{S,T,E_r\}$. Besides, one more $R$-vector ${\bf R_1}$  and a
scalar $A$ operators should be included in the set of group
generators. These generators are expressed in terms of Hubbard
operators as follows:
\begin{eqnarray}
&& R_1^+  = \sqrt{2}\left(X^{1E_r}-X^{E_r1}\right), R_1^-  =
\sqrt{2}\left(X^{E_r1}-X^{-1E_r}\right),
\nonumber\\
&& R_{1z}  =  -\left(X^{0E_r}+X^{E_r0}\right),\nonumber \\
&& A= i(X^{SE_r}-X^{E_rS}). \label{SE}
\end{eqnarray}
To close the algebra the commutation relations (\ref{comm1}) which
are valid also for $R_{1\alpha}$ should be completed by
\begin{eqnarray}
&& [R_{l\alpha},R_{1\beta}]=i\delta_{\alpha\beta}A,
[{R}_{1\alpha},A] = iR_{l\alpha},
\label{comm2} \\
&& [A,R_{l\alpha}]=iR_{1\alpha},\; \;\;[A,S_{l\alpha}]=0.
\nonumber
\end{eqnarray}
The system of commutation relations (\ref{comm1}), (\ref{comm2})
is that of the $o_5$ algebra, and the manifold $\{S,T,E_r\}$ obeys
$SO(5)$ dynamical symmetry, provided all three levels are involved
in the interaction within the Hamiltonian {\cal H}. The Casimir
operator for the $SO(5)$ group in this representation is ${\bf
S}^2 + {\bf R}^2 + {\bf R_1}^2 +A^2=4$.

In terms of these operators ${\cal H}_0$ has the form
\begin{equation}
{\cal H}_0=\frac{1}{2}\left(E_T {\bf S}^2 + E_S {\bf R}^2\right)+
Q(\hat{N}-2)^2. \label{1.3a}
\end{equation}
and
\begin{equation}
{\cal H}_0=\frac{1}{2}\left(E_T {\bf S}^2 + E_S {\bf R}^2 +
E_{E_r} {\bf R}_1^2 \right) +Q(\hat{N}-2)^2. \label{1.3b}
\end{equation}
for the $SO(4)$ and the $SO(5)$ groups, respectively. The last
terms in (\ref{1.3a}) and (\ref{1.3b}) control the number of
electrons given by the operator $\hat{N}$ in the DQD.

As is seen from this equation, spin is still conserved in the
isolated DQD. However, a tunnel contact with metallic leads breaks
the spin rotation invariance and reveals the dynamical symmetry of
the DQD. The physical mechanism of this symmetry breaking is {\it
electron cotunneling} with spin flips, when an electron with spin
$\sigma$ enters the DQD, whereas another electron with spin
$\sigma'$ leaves it. This process is known to be a source of Kondo
effect in tunnel barriers and quantum dots \cite{GLA}. Eliminating
charge degrees of freedom by means of the Schrieffer-Wolff
transformation, one usually arrives at an exchange-like
cotunneling Hamiltonian of the type $J_{cot}{\bf S}\cdot {\bf s}$,
where $J_{cot}\sim W^2$, and $W$ is a lead-dot tunneling amplitude
(which is anti-ferromagnetic, $J_{cot}>0$) .

Since $E_T-E_S=\delta>0$, the Kondo effect seems to be irrelevant
in DQD with even occupation. However, one should remember that the
tunneling $W$ induces additional contribution of indirect exchange
between the two wells of the DQD. As is shown in Refs. \cite{KA01}
this contribution may change the sign of $\delta$ provided the
excitation $E_r$ is soft enough, but the condition $V/(E_T-E_S)\ll
1$ is still valid. Then the exciton $E_r$ is eliminated from the
manifold, the symmetry of the DQD reduces from $SO(5)$ to $SO(4)$
and the Schrieffer-Wolff transformation yields the effective spin
Hamiltonian
\begin{equation}\label{1.10}
{\cal H}={\cal H}_0 + J_{cot}^T{\bf S}\cdot {\bf s}+
J_{cot}^{ST}{\bf R}\cdot {\bf s},
\end{equation}
where $J_{cot}^{T}$ and $J_{cot}^{ST}$ are two indirect exchange
coupling parameters which are renormalized by Kondo screening.
This screening affects both vectors ${\bf S}$ and ${\bf R}$.

The problem of Kondo tunneling within the framework of the
Hamiltonian (\ref{1.10}) has been solved in the weak coupling
limit (see \cite{KA01} and references therein), so we do not
elaborate upon it here. For further progress it is important to
note that this example demonstrates how the dynamical symmetry is
realized in the S/T subspace due to interaction with the electrons
in the leads, which breaks the rotational symmetry of the isolated
spin system. This interaction introduces its own energy scale
${\cal E}$ into the problem (in example considered above it is
just the Kondo temperature $T_K$), and the dynamical symmetry of
the DQD as a spin rotator becomes relevant when the T/S energy
splitting is comparable with $T_K$. In more complicated quantum
dots the spin manifolds consist of several S/T pairs, and the
dynamical symmetry of such dots is described by the $SO(n)$ groups
(see Ref. \cite{KKA} where the cases of $n=3,5,7$ are discussed).
\section{From spin rotator to spin ladder}
In this section we show that an $SO(4)$ symmetry is an intrinsic
property of S=1/2 spin ladders and decorated spin chains (defined
below) . A generic Hamiltonian for the spin systems under
consideration is of the Heisenberg-type, consisting of spin 1/2
moments residing on the sites of a two leg ladder,
\begin{equation}
H^{(SL)} =J_t\sum_{\langle i1,i2\rangle} {\bf s}_{i1}\cdot{\bf
s}_{i2} + J_l\sum_{\alpha}\sum_{\langle i\alpha,j\alpha\rangle}
{\bf s}_{i\alpha}\cdot{\bf s}_{j\alpha}. \label{spl}
\end{equation}
Here, the index $\alpha=1,2$  enumerates the legs of the ladder,
and the sites $\langle i1,i2\rangle$ belong to the same rung
(Fig.2a).

A chain of dimers of localized spins illustrated by Fig. 2b is
described by the simplified version of this Hamiltonian
\begin{equation}
H^{SRC}= J_t\sum_{\langle i1,i2\rangle}{\bf s}_{i1}\cdot{\bf
s}_{i2} +J_l\sum_{\langle ij\rangle}{\bf s}_{i1}\cdot{\bf s}_{j1}
\label{1}
\end{equation}
The geometry of alternate rungs is chosen in a system (\ref{1}) to
avoid exchange interaction between spins ${\bf s}_{i2}$ and ${\bf
s}_{j2}$. The transverse coupling may emerge either from direct
exchange (in case of localized spins) or from indirect
Anderson-type exchange induced by tunneling (similarly to the case
encountered in QD). In the latter case the sign of $J_t$ is
antiferromagnetic (AFM), in the former case it may be
ferromagnetic (FM) as well. The same is valid for $J_l$.
\begin{figure}[ht]
\centerline{\epsfig{figure=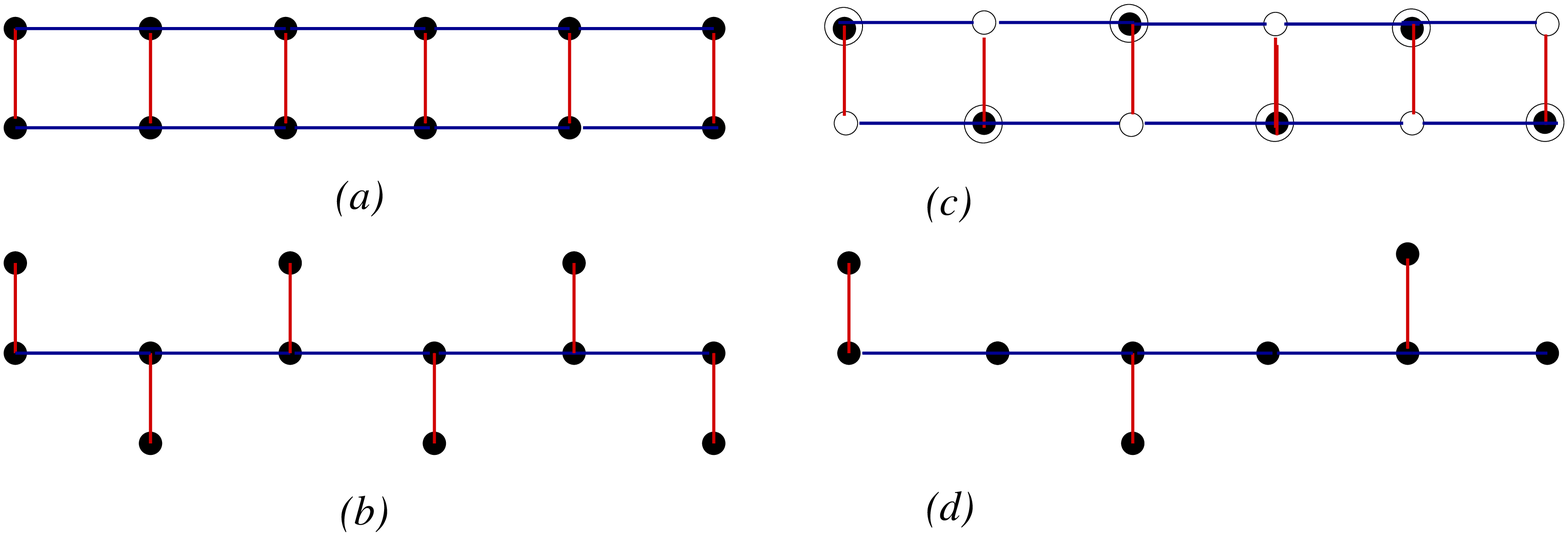,width=70mm,height=40mm,angle=0}}
\caption{Spin Ladder (a),  Spin Rotator Chain (b), Spin ladder in
the CDW phase (c) and Alternate Spin Rotator Chain (d)}
\label{fig2}
\end{figure}

It is useful to start with diagonalization of the Hamiltonian of a
perpendicularly aligned dimer (cf. Ref. \cite{Barn}). The $SO(4)$
symmetry stems from the obvious fact that the  spin spectrum of a
dimer $\{i1, i2\}$ is formed by the same singlet-triplet (ST) pair
as the spin spectrum of DQD studied in the previous section. This
analogy prompts us a canonical transformation connecting two pairs
of spin vectors, $\{{\bf s}_{i1},{\bf s}_{i2}\}$ and $\{{\bf
S}_{i},{\bf R}_{i}\}$: The two sets of spin operators are
connected by a simple rotation
\begin{equation}
{\bf s}_{i1}=\frac{{\bf S}_{i}+{\bf R}_{i}}{2},~~~ {\bf
s}_{i2}=\frac{{\bf S}_{i}-{\bf R}_{i}}{2}, \label{rot}
\end{equation}
Then the Hamiltonian ${\cal H}_i$ of a single dimer $i$ is the
same as the Hamiltonian (\ref{1.3a}) of DQD. The total spin of a
dimer is not conserved in such a spin chain, so the dynamical
symmetry of an individual rung is revealed by the modes
propagating along the chain \cite{Barn}. Applying the rotation
operation (\ref{rot}) to the Hamiltonians (\ref{spl}) and
(\ref{1}), we transform them into an equivalent form
\begin{equation}
{\cal H}={\cal H}_0 +{\cal H}_{int}. \label{1.1}
\end{equation}
Here ${\cal H}_{0}=\sum_{i}{\cal H}_{i}$ is common for both
models. It is useful to include  the Zeeman term in ${\cal H}_i$,
\begin{equation}
{\cal H}_i=\frac{1}{2}\left(E_S {\bf R}_i^2 + E_T {\bf
S}_i^2\right) + hS_{iz}. \label{1.3}
\end{equation}
We confine ourselves by a charge sector $N_i=2$ and omit the
Coulomb blockade term for the sake of brevity. The interaction
part of the SL Hamiltonian transforms under rotation (\ref{rot})
to the following expression
\begin{equation}
{\cal H}_{int}^{SL}= \frac{1}{4}J_l\sum_{\langle ij\rangle}({\bf
S}_i{\bf S}_j+{\bf R}_i{\bf R}_j) \label{I.2a}
\end{equation}
The interaction part of the SRC Hamiltonian is
\begin{equation}
{\cal H}_{int}^{SRC}= \frac{1}{4}J_l\sum_{\langle ij\rangle}({\bf
S}_i{\bf S}_j+2{\bf R}_i{\bf S}_j+{\bf R}_i{\bf R}_j) \label{I.2}
\end{equation}
One  may also consider the {\it alternate SRC model} (ASRC, see
Fig. \ref{fig2}(c)). After an appropriate rotation (\ref{rot})its
interaction Hamiltonian acquires  the form
\begin{equation}
{\cal H}_{int}^{ASRC}=\frac{1}{4}J_l\sum_{\langle ij\rangle}({\bf
S}_i{\bf S}_j+ {\bf S}_i{\bf R}_j). \label{I.5}
\end{equation}
Now we see that all three effective Hamiltonians belong to the
same family. In all  cases the initial ladder or "semi-ladder"
Hamiltonian is transformed into a one-dimensional spin-chain
Hamiltonian, which, however, takes into account the hidden
symmetry of a dimer. The effective Hamiltonians (\ref{I.2a}),
(\ref{I.2}), (\ref{I.5}) contain operators ${\bf R}$ describing
the dynamical symmetry of the dimers. This dynamical symmetry
turns the spectrum of this Hamiltonians to be richer than that of
a standard Heisenberg chain. Like in many other cases, rotation
transformation eliminates the antisymmetric combination of two
generators.

Thus, the transformation (\ref{rot}) reveals the hidden symmetry
of a spin 1/2 ladder (\ref{I.2a}). It maps the Hamiltonian onto a
pair of coupled chain Hamiltonians: one is the conventional spin 1
chain, while the other is a pseudospin chain. A spin ${\bf S}_i$
and a pseudospin ${\bf R}_i$ are kinematically coupled by the
commutation relations and by the local Casimir constraint
\begin{equation}
{\bf S}_i^2+ {\bf R}_i^2 =3. \label{cas}
\end{equation}

One may also compare the Hamiltonian (\ref{I.2a}) with the
effective Hamiltonian of a spin 1 chain, which arises after
decomposition of spin-one operators into a pair of spin 1/2
operators, ${\bf S}_i= {\bf s}_i + {\bf r}_i$ \cite{Luth}. This
decomposition operation transforms the initial Hamiltonian into a
form similar to $H^{SRC}$ but for spin-one-half operators ${\bf
s}_i,~{\bf r}_i$. The difference between the two cases is that
these effective spins commute, (unlike operators ${\bf S}_i,~{\bf
R}_i$). In other words, the difference is that the local symmetry
of spin-one chain is $SO(3)$ whereas the local symmetry of SRC is
$SO(4)$. The spin rotator chains (\ref{I.2}), (\ref{I.5}) are in
some sense intermediate between spin chains and spin ladders. In
these cases the spin-pseudospin symmetry is obviously broken by
the cross terms $2{\bf S}_i{\bf R}_j$.

The excitation spectrum of spin ladders may be calculated in terms
of operators ${\bf S}_i$ and ${\bf R}_i$. For example, the well
known expression for a gap $\Delta E$ in excitation spectrum in
the limit of strong transversal exchange $J_t \gg J_l$ for AFM
interaction \cite{Barn} looks like
\begin{equation}
\Delta E=J_t+ \frac{(J_l/4)^2 \sum_{ij,\alpha\beta}\left(\langle
T_{ij}^{\alpha\beta}|{\bf R}_i{\bf R}_j|S_iS_j\right)^2} {(E_T
-E_S)}=J_t+\frac{3J_l^2}{8J_t}
\end{equation}
(here $T_{ij}^{\alpha\beta}$  and $S_iS_j$ stand for possible
triplet projections and spin states at the sites $i,j$,
respectively). The singlet-triplet excitations above this gap are
given by the dispersion law $\omega(k)=\Delta E+ J_l \cos k$.

In all cases the simplified versions of Heisenberg Hamiltonians
may be considered. The simplified SL models are well known
\cite{Dag}. The anisotropic versions of the Hamiltonian
(\ref{I.2}) are: {\it Ising-like SRC model:}
\begin{equation}
H=\frac{1}{4}J_l\sum_{\langle
ij\rangle}(S^z_iS^z_j+2S^z_iR^z_j+R^z_iR^z_j). \label{I.1}
\end{equation}
{\it Anisotropic SRC model:}
\begin{equation}
H=\frac{1}{4}J_l\sum_{\langle
ij\rangle}\left[(S^+_iS^-_j+S^+_iR^-_j+S^-_iR^+_j+R^+_iR^-_j)\right.
\label{I.3}
\end{equation}
$$
+\left.\Delta(S^z_iS^z_j+2S^z_iR^z_j+R^z_iR^z_j)\right].
$$

{\it SRC in strong magnetic field:} SO(4) group reduces to SU(2)
group in magnetic field, when the Zeeman splitting {\it exactly}
compensates the exchange gap in a single dimer, $h_0=|E_T-E_S|$.
Then at low T, the states $|i0\rangle$ and $|i-1\rangle$ are
quenched, and only two components, $R^\pm$ survive in the manifold
(\ref{m.1}), (\ref{SP}). As a result, the Hamiltonian (\ref{I.2})
is mapped onto a $XY$-model for spin 1/2:
\begin{equation}
H_{XY}^{(R)}= \frac{1}{4}J_l\sum_{\langle ij\rangle}(R_i^+R_j^- +
H.c.). \label{I.4}
\end{equation}
This means that starting from a singlet ground state for
$J_t\equiv E_T-E_S>0$, one may induce development of spin
liquid-like excitations by applying strong magnetic field. In a
near vicinity of this point of degeneracy, $H_{int}$ acquire the
features of {\it XY model in transverse magnetic field}.

\section{Fermionization}
To describe the elementary excitations in SRC, one should
generalize the $SU(2)$-like semi-fermionic representation for $S$
operators \cite{popov}
$$
S^+  =   \sqrt{2}(f_0^\dagger f_{\bar{1}}+f^\dagger_{1}f_0),\;\;\;
S^-  =  \sqrt{2}(f^\dagger_{\bar{1}}f_0+ f_0^\dagger f_{1}),
$$
\begin{equation}
S_z  =  f^\dagger_{1}f_{1}-f^\dagger_{\bar{1}}f_{\bar{1}},
\label{spin}
\end{equation}
where $f^\dagger_{1}$, $f^\dagger_{\bar{1}}$ denote  creation
operators for fermions with spin ``up'' and ``down'' respectively
whereas $f_0$ stands for spinless fermion \cite{popov,kis}.
Fermionization of  $SO(4)$ group is completed by introducing one
more spinless fermion $f_s$ which represents the singlet state. As
a result, $P$-operators are given by the following equations:
$$
P^+ =  \sqrt{2}(f^\dagger_{1} f_s -  f_s^\dagger
f_{\bar{1}}),\;\;\; P^- =  \sqrt{2}(f_s^\dagger f_{1} -
f^\dagger_{\bar{1}}f_s),
$$
\begin{equation}
P^z  =  -( f_0^\dagger f_s + f_s^\dagger f_0).
 \label{proj}
\end{equation}
Then the single-site Hamiltonians may be represented in a form
\begin{equation}
H_i=-\delta f_{is}^\dagger f_{is}+
h(f^\dagger_{i1}f_{i1}-f^\dagger_{i\bar{1}}f_{i\bar{1}})
\label{hi}
\end{equation}
The  Casimir operator ${\bf S}^2+{\bf P}^2=3$ transforms to the
local constraint
$$\sum_{\Lambda=\pm,0,s}f^\dagger_\Lambda f_\Lambda=1.$$
We start the studies of elementary excitations in SRC with the
anisotropic XXZ version. of general effective Hamiltonian. The
simplest of all is the case (\ref{I.5}). The problem is reduced to
a standard XY-model for spin one half, and the spinon spectrum may
be easily obtained either by bosonization  or by spinon-type
fermionisation. In former case one deals with hard-core bosons,
and in latter one the problem is mapped onto the non-interacting
incompressible fermions at half-filling.

We concentrate on a more complicated case of XXZ-SRC model
(\ref{I.3}) specifically on its simplified alternate version,
which is obtained from the Hamiltonian (\ref{I.5}). The
Hamiltonian of this model is
\begin{equation}
H=\frac{1}{4}J_l\sum_{\langle
ij\rangle}(S^+_iS^-_j+S^+_iP^-_j+S^-_iP^+_j +\Delta( S^z_iS^z_j+
2S^z_iP^z_j).
 \label{I.6}
\end{equation}
The S-S part of this Hamiltonian describes the S=1 chain, with the
Haldane gap in the excitation spectrum (see,
e.g.,\cite{batort,halgap}). The question is, how do the S-P
interaction modifies the gap. We consider the case of FM dimers,
when the triplet is the ground state. In this case one has one
more gap mode, where the gap equals $J_t$. This mode is coupled to
Haldane branch only via S-P exchange terms in (\ref{I.6}).

The spin liquid fermionization approach adopted here is a
convenient tool for description of Haldane spectrum. Unlike the
S=1/2 model, where the spin-liquid state is easily described by
globally U(1) invariant modes
$T_{ij}T_{ji}=\sum_{\sigma}f^\dagger_{i\sigma}f_{j\sigma}|^2,$ in
case of S=1, one deals with variables which effectively break this
symmetry. One can rewrite the effective Hamiltonian of SRC model
with $\Delta=0$ in a form
\begin{eqnarray}
&&H=\frac{1}{4}J_l\sum_{ij}\left(f^\dagger_{i1}f_{j1}+
f^\dagger_{i\bar{1}}f_{j\bar{1}}\right) \bar
B_j^{0S}B_i^{0S}\nonumber\\
&&+f^\dagger_{i\bar{1}}f^\dagger_{j1} C_j^{0S}B_i^{0S}+ \bar
B_i^{0S}\bar C_j^{0S}f_{j1}f_{i\bar{1}}, \label{anom}
\end{eqnarray}
where $B_j^{0S}=f_{0j}+f_{Sj}$ , $C_j^{0S}=f_{0j}-f_{Sj}$. The
terms in the first line of Eq. (\ref{anom}) describe coherent
propagation of spin fermions accompanied by a backflow on neutral
fermions, whereas the terms in the second line are "anomalous"
(they do not conserve spin fermion number).
 For example the propagator $\langle S^+_iS^-_j\rangle$
contains anomalous components $
f^\dagger_{i1}f^\dagger_{j\bar{1}}f_{j0}f_{i0} \to
F^*_{ij,1\bar{1}}F_{ji,00}$  along with normal ones
$f^\dagger_{i1}f_{j1}f^\dagger_{j0}f_{i0}.$ Here
$F_{ij,\Lambda\Lambda'}=f_{j\Lambda}f_{i\Lambda'}$. The first term
in (\ref{anom}) describes the kinetic energy spinon excitations,
and two last anomalous term breaking U(1) symmetry are responsible
for the Haldane gap. To reveal the contribution of dynamical
symmetry on the Haldane gap, one have to note that the terms
$B^{0S}$ and $B^{0S}$ appear both as a counterflow in the first
term and as gauge symmetry breaking terms in the second line. In
spin 1 ladder the counterflow term $\sim f^\dagger_{i0}f_{j0}$
predetermines the width of spinon band described by the first line
of Eq. (\ref{anom}). Apparently, the one more channel
(tripet/singlet transitions in $B^{0S}$) enhances this effect,
because in this case the local constraint imposes more
restrictions of phase fluctuations.
 The gap itself is due to
anomalous correlations described by the second line of  Eq.
(\ref{anom}). Here the appearance of second channel of spinless
excitations results in formation of even and odd operators
$B_j^{0S}$ and $C_j^{0S}$. The Haldane gap closes when the $|0
\rangle$ and $|S \rangle$ states are degenerate (the odd operator
$C_j^{0S}$ nullifies the anomalous terms responsible for its
formation). This means that appearance of $0S$ channel {\it
favors} closing of the Haldane gap.

\begin{figure}[ht]
\centerline{\epsfig{figure=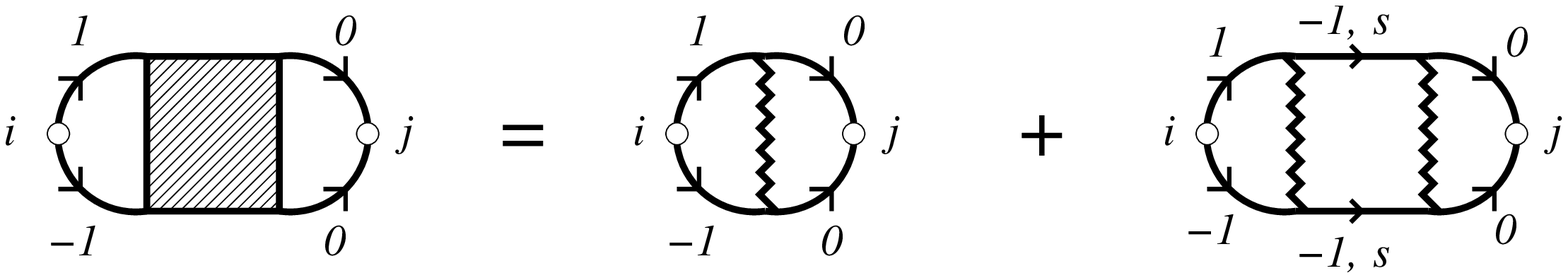,width=90mm,height=20mm,angle=0}}
\caption{Lowest order contributions to anomalous propagator.}
\label{fig3}
\end{figure}

In a strong coupling  case of $J_t\gg J_l$ both above trends may
be considered at least in the lowest order of a perturbation
theory. In case of spin ladders \cite{Barn} the 1-st and
2-nd-order  in $g=J_l/J_t$ anomalous diagrams are represented in
Fig.\ref{fig3}.
\section{SO(n) dynamical symmetries for a two-leg quantum ladder}
It was mentioned in Section II that the dynamical symmetry of DQD
becomes $SO(5)$, if the charge transfer excitonic state is
involved (see Eq. \ref{1.3b}). In this section we discuss the
origin of this symmetry in spin ladders. This problem arose in a
context of $SO(5)$ symmetric $t-J$ model of 2D cuprate
superconductors \cite{Zhang}. Later on the version of this theory
was formulated for cuprate  two-leg ladders \cite{Hanke}. Here we
show that the {\it dynamical} $SO(5)$ group  arises in description
of Heisenberg ladder, but excitonic states are involved in this
symmetry instead of Cooper states.

Let us consider a two-leg quantum ladder depicted in Fig.2a under
condition of strong Coulomb blockade imposed  on each rung $i$. We
allow electron tunneling $t_{ij}^\alpha$ along both legs. This
tunneling is described by the Hamiltonian
\begin{equation}\label{5.1}
{\cal H}_{tun}=\sum_{ij}\sum_{\alpha\sigma} t^\alpha
d^\dag_{i\alpha\sigma} d_{j\alpha\sigma}.
\end{equation}
(only a nearest-neighbor hopping along the leg is allowed). This
hopping results in appearance of {\it charged} rungs because each
hopping act creates a hole on a rung $j$ and an electron on a rung
$i$. To treat this charging properly the Coulomb blockade term in
the Hamiltonian ${\cal H}_i$ (\ref{1.3}) should be restored (see
Eq. \ref{1.3b}), and the terms with excess electron and excess
hole should be added. It is more convenient to represent the
Hamiltonian ${\cal H}_i$ of individual rung $i$ in terms of
diagonal Hubbard operators [see (\ref{HX})]
\begin{eqnarray}
{\cal H}_i=\sum_i\left[\sum_{\Lambda}E_{\Lambda}
X_i^{\Lambda\Lambda}+\sum_{\gamma}E_{\gamma} X_i^{\gamma\gamma}+
\sum_{\Gamma}E_{\Gamma} X_i^{\Gamma\Gamma}\right]
\end{eqnarray}\label{5.2}
Here index $\gamma=\alpha\sigma$ stands for the states with one
electron with spin $\sigma$ on a site $i\alpha$ of the rung $i$,
index $\Gamma=\alpha\sigma$ stands for three-electron states of a
rung, where two electrons occupy site $i\alpha$ and one electron
with spin $\sigma$ is located in a site $i\bar{\alpha}$
($\bar{\alpha}=2$ if $\alpha=1$ and v.v). The energy levels
$E_\gamma$ and $E_\Gamma$ are separated by a Coulomb blockade gap
$\sim Q$ from the two-electron states $E_\Lambda$. The Hamiltonian
(\ref{5.1}) in these terms is
\begin{equation}\label{5.3}
{\cal H}_{tun}=\sum_{ij,\alpha}\sum_{\gamma\Gamma\Lambda} t^\alpha
X_i^{\Gamma\Lambda}X_j^{\gamma\Lambda}+ H.c.
\end{equation}

It is seen from (\ref{5.3}) that the intersite hopping "charges"
two neighboring rungs in a ladder, which was initially neutral,
and one should pay the energy $\sim Q$ for each hopping act, like
in the generic Hubbard model at half-filling. This energy loss is
reduced if an electron-hole pare is created at a given rung $i$.
In this case the electron-hole attraction $V<0$ partially
compensates charging energy $Q$. Let us assume the hierarchy $Q\gg
Q-|V| \gg t$. Then the states $|\Gamma\rangle$ may be excluded
from the manifold in favor of excitonic states $|iE_\alpha\rangle$
similar to the states $|E_r\rangle$ introduced in (\ref{SE}). Here
$\alpha=1$(2) for the electron occupying  site $i1 (i2)$. If the
ground state of a rung is singlet, $|iS\rangle$, then electron and
hole have antiparallel spins and the excitation energy is
$Q'=Q-|V|$.
 Even combination of two states $|iE_{(1,2)}\rangle$
form a singlet exciton $|iE\rangle$. Such exciton can propagate
coherently along the ladder unlike single electron, whose
tunneling leaves a trace of charged states according to
(\ref{5.3}). Indeed, translation of e-h pair from a rung $i$ to a
neighboring rung $i+1$ can be presented as coherent tunneling of
electron from a site $i\alpha$ to a site $i+1,\alpha$ and another
electron in the opposite direction (from $i+1, \bar{\alpha}$ to
$i, \bar{\alpha}$. The exciton propagation is described by the
following term in effective Hamiltonian:
\begin{equation}\label{5.4}
H_{ex}=\sum_{i}K^{S}X_i^{SE}X_{i\pm 1}^{ES}
\end{equation}
with effective exchange coupling constant $K^S=|t_1t_2|/Q'$, and
the dispersion law describing coherent exciton propagation is
$\epsilon_S(k)=2K^S\cos k$. As was shown in Section 2, the
manifold $\{iS,iT,iE\}$ possesses the local dynamical symmetry
$SO(5)$ [see Eqs. (\ref{SE}), (\ref{comm2})], and this symmetry
allows existence of coherent collective singlet exciton mode. The
Hamiltonian (\ref{5.4}) acquires a form
$H_{ex}=(K^{S}/4)\sum_{ij}\widetilde{A}_i\widetilde{A}_j$ in terms
of generators of SO(5) group (\ref{SE}), where
$\widetilde{A}_i=({\bf R}_i^2-1)A_i$. There is one more collective
mode, namely triplet exciton $|E_\mu\rangle$ ($\mu=\pm1,0$)
separated by the gap $\sim J_t$ from the singlet exciton. In case
of triplet ground state ($J_t <0$), this mode becomes the lowest
one, and the Hamiltonian similar to (\ref{5.4}) may be derived for
triplet exciton propagation with operators $X_i^{TE_\mu}$
replacing $X_i^{SE}$. In this case the manifold $\{iS,iT,iE_\mu\}$
consists of one singlet and two triplets, and the corresponding
dynamical group is $SO(7)$ \cite{KKA}. If exchange and excitonic
gaps are comparable in magnitude, then the interplay between
exciton and magnon modes is possible, and dynamical symmetry will
result in observable physical effects. Like in cuprate ladder,
\cite{TF}, the excitonic instability can develop for certain
values of model parameters, which results in phase separation and,
in particular in formation of CDW phase illustrated by Fig.
\ref{fig1}c (where double and empty circles stand for doubly
occupied and empty sites respectively.
\section{Concluding remarks}
We rederived a family of Hamiltonians for quantum dots and quantum
ladders in terms of $SO(4)$ group, which describes the dynamical
symmetry of spin rotator \cite{KA01}. We exploited the fact that
in case, when the Hamiltonian ${\cal H}$ contains blocks ${\cal
H}_i$ formed by two sites occupied by spins 1/2, one may use its
eigenstates (singlet-triplet manifolds) as a basis for
representing the spin invariants entering ${\cal H}$. These
invariants contain the Runge-Lenz-like vectors ${\bf R}_i$ along
with the usual spin vectors ${\bf S}_i$. If the electron-hole
pairs are also included in the set of eigenstates, then the local
dynamical symmetry of ${\cal H}_i$ is characterized by the $SO(n)$
group with $n=5$ or 7 for a singlet and triplet ground state of
${\cal H}_i$, respectively. The elementary excitations in quantum
dots and quantum ladders are described by means of generators of
$SO(n)$ groups and the interplay between different branches of
excitation spectra is a direct manifestation of local dynamical
symmetry violated by non-local interactions.

\end{document}